\documentclass[a4paper, 10 pt]{article}
\title{ Learning Political DNA in the  Italian Senate}

\usepackage[latin1]{inputenc}
\usepackage{graphicx,subfig,float,dblfloatfix}
\usepackage{amssymb,amsmath,dsfont}
\usepackage{url}
\usepackage{layaureo}
\usepackage{wrapfig,color}
\usepackage{pgfplots}
\usepackage{xcolor}
\usepackage{color, colortbl}

\newcommand{\R}{\mathbb{R}} 
\newcommand{\N}{\mathbb{N}}  

\def\Prob{\mathbb{P}}

\newcommand{\Ncal}{\mathcal{N}}

\newcommand{\norm}[1]{\|#1\|}
\newcommand{\abs}[1]{|#1|}

\definecolor{LightGray}{gray}{0.96} 
\definecolor{DarkGray}{gray}{0.8}
\author{
Antonio Longo\thanks{A.~Longo is a MSc student at
        Politecnico di Torino, Corso Duca degli Abruzzi, 10129 Torino, Italy.}, Chiara Ravazzi\thanks{C. Ravazzi and F. Dabbene are with the National Research Council of Italy, CNR-IEIIT, c/o Politecnico di Torino, Corso Duca degli Abruzzi 14, 10129 Torino, Italy.},\\  Fabrizio Dabbene, Giuseppe Calafiore\thanks{G. Calafiore is with
        Politecnico di Torino (DET) and Research Associate of CNR-IEIIT, Corso Duca degli Abruzzi, 10129 Torino, Italy.}}

\usepackage{booktabs,colortbl, array}
\usepackage{pgfplotstable}
\pgfplotsset{compat=1.8}
\definecolor{rulecolor}{RGB}{0,71,171}
\definecolor{tableheadcolor}{gray}{0.92}
\newcommand{\topline}{ %
        \arrayrulecolor{rulecolor}\specialrule{0.1em}{\abovetopsep}{0pt}%
        \arrayrulecolor{tableheadcolor}\specialrule{\belowrulesep}{0pt}{0pt}%
        \arrayrulecolor{rulecolor}}
\newcommand{\midtopline}{ %
        \arrayrulecolor{tableheadcolor}\specialrule{\aboverulesep}{0pt}{0pt}%
        \arrayrulecolor{rulecolor}\specialrule{\lightrulewidth}{0pt}{0pt}%
        \arrayrulecolor{white}\specialrule{\belowrulesep}{0pt}{0pt}%
        \arrayrulecolor{rulecolor}}
\newcommand{\bottomline}{ %
        \arrayrulecolor{white}\specialrule{\aboverulesep}{0pt}{0pt}%
        \arrayrulecolor{rulecolor} %
        \specialrule{\heavyrulewidth}{0pt}{\belowbottomsep}}%

\pgfplotstableset{normal/.style ={%
        header=true,
        string type,
        font=\addfontfeature{Numbers={Monospaced}}\small,
        column type=l,
        every odd row/.style={
            before row=
        },
        every head row/.style={
            before row={\topline\rowcolor{tableheadcolor}},
            after row={\midtopline}
        },
        every last row/.style={
            after row=\bottomline
        },
        col sep=&,
        row sep=\\
    }
}

\begin{document}

\maketitle
\begin{abstract}
Motivated by the increasing interest of the control  community towards social sciences and the study of opinion formation and belief systems, in this paper we address the problem of
exploiting voting data for inferring the underlying affinity of individuals to competing ideology groups.

In particular, we mine key voting records of the Italian Senate during the XVII legislature, in order to extract the hidden information about the closeness of senators to political parties, based on a parsimonious feature extraction method that selects the most relevant bills.
Modeling the voting data as outcomes of a mixture of random variables and using sparse learning techniques, we cast the problem in a probabilistic framework and derive an information theoretic measure, which we refer to as Political \textit{Data-aNalytic Affinity} (Political DNA). The advantages of this new affinity measure are discussed in the paper.
The results of the numerical analysis on voting data unveil underlying relationships among political exponents of the Italian Senate.
\end{abstract}

\section{Introduction}
In the past decades, many efforts have been spent by the control community on the study of mathematical models for opinion formation in social and belief systems \cite{Friedkin2015ThePO}.
Among these models, Friedkin and Johnsen's (F\&J) opinion dynamics \cite{Friedkin:Johnsen:1999} has been experimentally validated for deliberative groups of small and medium size \cite{Friedkin:Johnsen:2011}. According to this model, the agents' opinions evolve as a convex combination of others' beliefs and an initial condition. In this sense, agents are not completely open-minded, being persistently driven by an individual attachment due, for example, to the influence from a specific ideology \cite{ca21d555c7db4f388ed00f8a07214c01}.
The key ingredient for  estimating this {\em stubborness} and, consequently, for offering insights in efficient control strategies for steering social behaviors towards desired patterns, is the development of new technically sound tools, able to extract low-dimensional information from social data \cite{618b8cb6d9a24441839f83010f95d1c2,Wai2016ActiveSO,Ravazzi2018}.

Following a  machine learning approach, in this paper we pose and solve the  problem of learning the attachment of individuals to their own community and the underlying influence from competing ideology groups using  observations of voting records.

For this purpose, the Italian political scenario represents a very interesting case of study for its complexity, compared to other foreign deliberative institutions, as it is composed by several parties.
Our data source is the nonprofit organization OpenPolis \cite{openpolis}, that tracks information about representative and senators in Italy, including votes, monitoring government daily events and providing statistics on politicians' behaviors.
In our analysis, we focus on data related to the activity in the Italian Senate during XVII legislature that are classified by  OpenPolis as {\em key votes}, i.e.\ considered as the most important both for the relevance of the subject matter and for the political value \cite{openpolis}. We also acquire the nominal membership of each senator to her/his political group, which will be used as side information.

The main contribution of this paper is the introduction of a new measure based on a features selection. This similarity index, which we refer to as Political Data-aNalytics Affinity (Political DNA), summarizes the degree of fidelity to the party and the influence from other groups. It can be equivalently interpreted as a quantitative indicator of the degree of rebellion to the {\em discipline} of the group. This definition is based on an information-theoretic ground, by modeling the votes as outcomes of a mixture of random variables and reformulating the computation of Political DNA as an estimation of class posterior probabilities. The combination of this new measure with sparse learning techniques allows us to select the most relevant bills determining variances in the political positions and to automatically identify outliers, i.e. senators that are not good representative of their nominal group.
Our analysis is directly interpretable and contributes to a better understanding of relationships and opinions in belief systems from a system and control point of view.

\subsection{Related works}
Several approaches have been proposed in the literature for scoring the political ideology from  voting data \cite{MS2006,poole1985spatial}.
The most popular techniques belong to the family of Multidimensional scaling, such as Nominate, W-Nominate, and DW-Nominate \cite{poole1985spatial,poole2000nonparametric}.
These methods are mainly used to produce graphical interpretations of political positions, representing
high-dimensional data in a space with a lower dimension.
In \cite{Jenkins2006} these methods have been applied to extract ideal points used as features in estimating party influence and to determine the ideological rank order in the US Congress \cite{Wai2016ActiveSO}.
Other works use these techniques to detect stubborn agents in opinion dynamics, assuming they should be the ones with extreme scores, i.e.\ who are either far left or far right in ideology \cite{fd9428313d5946bb87628aa2fe9b6f04}.

It is worth mentioning that our goal is to estimate not only the attachment of senators to their own ideology, but also to quantify the influence from other groups.
To estimate the similarity between two objects, several distances and similarity metrics can be used \cite{Weller2015}. However, in this paper, we are interested in a quantitative indicator vector, whose components naturally sum to one and can be interpreted as a probability of being influenced by a specific group.
This main feature can be used to mine social trust in F\&J model and is useful for representation of political positions as a convex combination of features, building what we call a \textit{Political Map}, a graphical tool for displaying multivariate data.

Besides Multidimensional scaling, many other techniques could in principle be applied to embed high dimensional data in a low dimensional space, such as Non-negative Matrix factorization \cite{calafiore2014optimization}, PCA \cite{jolliffe2011principal}, Sparse PCA \cite{ghaoui2007sparse}, to mention just a few. All these methods may depend on specific choices, e.g.\ encoding scheme, parameter selection and initialization, introducing some instability and making difficult a general comparison \cite{Brigadir2016}. However, we  show that the analysis of Political DNA based on Sparse PCA allows direct and better interpretation of the results, providing a list of the main bills determining the separation in the political positions of  senators, and an automatic method for identifying outliers.

\subsection{Outline}
The remainder of this paper is organized as follows. In Section \ref{sec:PF}, we give a brief overview of the available dataset and a description of the preprocessing procedure. Section~\ref{sec:DNA} addresses the problem of learning the political influence of senators from groups. The definition of Political DNA is formally introduced and the main  techniques for its computation and dimensionality reduction are discussed in detail. Numerical results are provided in Section~\ref{sec:NR}. Finally, some concluding remarks and discussions on future developments are reported in Section~\ref{sec:CR}.
The notation is stated next.

\subsection{Notation}
Natural and real numbers are denoted by $\N$ and $\R$, respectively. We denote column vectors with lower-case letters, and matrices with capital letters. {Given a matrix $X\in\R^{m\times n}$, $X^{\top}$ denotes its transpose, $X_{ij}$ is the entry corresponding to $i$-th row and $j$-th column. We use the notation $x^{(i)}\in \R^n$ for the column vector corresponding to the $i$-th row}, i.e.
$$
X=\left[\begin{array}{c}
{x^{(1)}}^{\top}\\
\vdots \\
{x^{(m)}}^{\top}
\end{array}\right].
$$
  We consider the following norms: given $z\in\R^n$, we denote the Euclidean norm and $\ell_0$-pseudonorm (number of non-zero elements) with $\|z\|_2$ and $\|z\|_0$, respectively. We denote the Frobenius norm of a matrix $X \in \R^{m \times n}$ as $\norm{X}_F$ where $\norm{X}_F := \sqrt{\sum_{i=1}^m\sum_{j=1}^n X_{ij}^2}$. Finally, we use the symbol  $\text{supp}\left(z\right)=\{i\in\{1,\ldots, n\}: z_i\neq0\}$ to denote the set of nonzero elements of $z$.
\section{Preliminaries}
\label{sec:PF}
\subsection{Dataset and problem statement}
The dataset contains the votes of the Italian Senate during the XVII legislature, which has been active from the 15th of March 2013 to the 22nd of March 2018.
Our analysis focuses on key final votes, spanning through the whole legislature across three governments.
The data contains a list of the senators. For each bill, identified by its title and a brief  description of the subject, the preference expressed by each senator is specified as: YES (\emph{``Favorevole''}), NO (\emph{``Contrario''}) and NV~(\emph{``Assente''}).
NV refer to cases in which the member was physically absent or was present but did not vote and did not take part in determining the {\em quorum} (unfortunately, the available data does not allow to distinguish among these cases).

Another field specifies the membership of the senators. In particular, we consider $n_g=7$ main groups: Partito Democratico (PD), Popolo della Libert��(PdL), Lega, Movimento 5 Stelle (M5S), Nuovo Centrodestra (NCD), Liberi e Uguali (LeU), and Other. The ``Other" class includes the mixed group (``{Gruppo Misto}'') as well as other minor political groups. In case a group is composed of multiple parties, only the most representative one has been indicated.
  For each senator the membership to the most recent political group has been considered.
   {{ The dataset is cleaned by removing senators that never voted on any bill,
   and all bills that were voted in a secret ballot. After this cleaning procedure the total number of senators is $m = 335$ and the total number of bills is $n = 155$.}}
Our goal is to infer a measure that describes the political influence
on each senator
from all groups, using the observations of the voting records.

\subsection{Data encoding and preprocessing}\label{sub:encoding}
The acquired data are collected in the vote matrix {$Z\in\{-1,0,+1\}^{m\times n}$}, where rows represent senators ($m=335$) and columns correspond to bills ($n=155$).
The entry $Z_{ij}$ contains a ternary value and  encodes the preference of senator $i$ on the specific bill $j$. More precisely, we assign $+1$ for YES, $-1$ for NO, and $0$ for NV.
{{Finally, we standardize the data over the columns, as in \cite{krzanowski2000principles,jackson2005user}:
$$		X_{ij} := \frac{Z_{ij} - \frac{1}{m}\sum_{i=1}^mZ_{ij}}{\sqrt{\sum_{i=1}^m(Z_{ij}- \frac{1}{m}\sum_{i=1}^mZ_{ij})^2}}.
$$
From now on we will work with the standardized matrix $X\in\R^{m\times n}$ and we will denote its rows by $\{x^{(i)\top}\}_{i=1}^m$ where $x^{(i)}\in\R^n$.}
}
{Vector $x^{(i)}\in\R^n$ thus contains the (standardized) votes of senator $i$ on the $n$ considered bills.}
\bigskip

Our problem can then be cast as follows: \textit{For each senator $s\in\{1,2,\ldots,m\}$, we want to infer the Political DNA, i.e. a vector $\pi^{(s)}\in[0,1]^{n_g}$ whose entries $\{\pi^{(s)}_g\}_{g\in\{1,\ldots,n_g\}}$ represent the influence of group $g$ to senator $s$, with the  property
that $
\sum_{g=1}^{n_g}\pi_g^{(s)}=1.
$}

\section{Inference of Political DNA}\label{sec:DNA}
{We reduce the problem to a soft discrimination task with $n_g$ classes (the political groups) and $m$ training data samples (the vote vectors of each senator). Given a set of data $\mathcal{D} = \{x^{(1)}, \ldots, x^{(m)}\}$ with  known class labels $\{\omega_1,  \ldots , \omega_m\}$, where $x^{(i)} \in \R^n$ and $\omega_i\in\{1,\ldots, n_g\}$, a generative probabilistic model is build for representing the data. }
Then, for each test vector we estimate the class posterior probabilities. We describe the procedure in detail assuming the data are described by Gaussian Mixtures Models (GMM). \subsection{GMM based discriminative anaysis}\label{sec:GMM}
\subsubsection{GMM}
The GMM assumes that the distribution of
the vote vector, given the class (i.e.,  the {\em class conditional density}), is a multivariate Gaussian distribution with mean $\mu_\ell\in\R^n$ and covariance matrix $\Sigma_{\ell}$, for each class $\ell\in\{1,\ldots, n_g\} $.
The resulting (unconditional) distribution of the vote vector is a finite mixture of normal distributions, with $n_g$ components, that is
\begin{equation}
		f(x) = \sum_{\ell = 1}^{n_g} \alpha_{\ell} f_{\ell}(x;\mu_{\ell},\Sigma_{\ell}),
\end{equation}
where $\alpha_\ell\geq 0$,  $\sum_{l = 1}^g \alpha_{\ell} = 1$, are the prior
mixing probabilities,
 and $f_{\ell}(x)$ is the normal probability density function associated to the $\ell$-th political group:
\begin{equation}
		f_{\ell}(x) = \frac{\exp\left(-\frac{1}{2}\left(x - \mu_{\ell}\right)^\top \, \Sigma_{\ell}^{-1} \, \left(x - \mu_{\ell}\right)  \right)}{\sqrt{\left(2\pi\right)^{n} \, \mathrm{det}\left(\Sigma_{\ell}\right)}},
\end{equation}
$\mu_{\ell} \in \R^{n}$ and $\Sigma_{\ell} \in \R^{n\times n}$ are the mean vector and the covariance matrix associated to the $\ell$-th group.
 Equivalently,
\begin{equation}
x^{(i)} \sim \begin{cases}
		  \Ncal\left(\mu_1,\Sigma_1\right), & \mbox{with probability } \alpha_1 \\
		  \vdots \\
		  \Ncal\left(\mu_{n_g},\Sigma_{n_g}\right), & \mbox{with probability } \alpha_{n_g}.
		  \end{cases}
\end{equation}
GMM is fully characterized by $\{\alpha_{\ell},\mu_{\ell},\Sigma_{\ell}\}_{\ell\in\{1,\ldots,n_g\}}$.
For each group, the class-conditional  mean and the covariance matrix are computed via maximum likelihood (ML)  \cite{MARIN2005459}.

Defining $G_\ell=\{i:\omega_i = \ell\}$, we have:
\begin{gather*}	\alpha_{\ell} = \frac{\abs{G_\ell}}{m}	,
\quad
	\mu_{\ell} = \frac{1}{\abs{G_\ell}} \, \sum_{i \in G_\ell} x^{(i)},   \\
	\Sigma_{\ell} = \frac{1}{\abs{G_\ell} - 1} \sum_{i \in G_\ell}	\, \left(x^{(i)}   - \mu_{\ell}\right)^{\top} \, \left(x^{(i)}   - \mu_{\ell}\right),
	\end{gather*}
where with $\abs{G_\ell}$ we indicate the cardinality of $G_\ell$, i.e. the number of Senators who nominally belong to the $\ell$-th group.


	\subsubsection{Posterior distribution evaluation}
Once the model is characterized as described above, for each datum $i$ we introduce the hidden variable $z_{i\ell}$, such that  $z_{i\ell}=1$ if the test data $x^{(i)}$ comes from class $\ell\in\{1,\ldots, n_g\}$, and $0$ otherwise.
The Political DNA is then given by the
posterior  probabilities $\pi^{(i)}_{\ell}$
\begin{align}\begin{split}\label{eq:pi}
	\pi^{(i)}_{\ell} &= \Prob\left(z_{i\ell} = 1 \mid x^{(i)};\{\alpha_{\ell}, \mu_{\ell}, \Sigma_{\ell}\}_{\ell\in\{1,\ldots, n_g\}} \right)
	\end{split}
\end{align}
which represent the posterior belief that senator $i$ belongs to group $\ell$, based on the GM model and given the evidence
	provided by the vote vector $x^{(i)}$. By applying Bayes' theorem, we compute
	the desired quantities as
\begin{align}\begin{split}\label{eq:pi}
	\pi^{(i)}_{\ell}
	&= \frac{\alpha_{\ell} \, \frac{\exp\left(-\frac{1}{2}\left(x^{(i)} - \mu_{\ell}\right)^\top \, \Sigma_{\ell}^{-1} \, \left(x^{(i)} - \mu_{\ell}\right)  \right)}{\sqrt{\mathrm{det}\left(\Sigma_{\ell}\right)}}
	}{\sum_{j = 1}^{n_g} \alpha_j \frac{\exp\left(-\frac{1}{2}\left(x^{(i)} - \mu_{j}\right)^\top \, \Sigma_{j}^{-1} \, \left(x^{(i)} - \mu_{j}\right)  \right)}{\sqrt{\mathrm{det}\left(\Sigma_{j}\right)}}}.
		\end{split}
\end{align}

The method depends heavily on the class covariance estimation.
In our specific application, we can encounter two problems:
\begin{enumerate}
\item if the class has only a few examples, as in small parties, the class covariance may suffer in accuracy.
\item to compute the posterior distribution in \eqref{eq:pi} the inversion of the covariance matrix is needed. When we have  a large number of variables and few observations, the covariance matrix might be not of full rank.
\end{enumerate}To avoid these problems we map the data into a lower-dimensional space.

\subsection{Dimensionality reduction: PCA vs Sparse PCA}\label{sub:DR}
To increase the insight on the data and to better expose the political information embedded in the vote matrix, we project the data on a $k$-dimensional subspace (with $k \leq n$). In particular, we apply PCA \cite{jolliffe2011principal} and Sparse PCA \cite{ ghaoui2007sparse}, whose main principles are briefly explained next.

\subsubsection{PCA}
Principal component analysis (PCA) is a widely known technique used to find the most important directions in a dataset. These are orthonormal directions along which the data has the most variation and are called the PCs. The PCs  induce a new set of variables ordered in such a way that the first few of them will retain the most amount of variation. Thus, by discarding all but the first few directions the data can be modelled along a low-dimensional subspace exposing the most relevant components of the information.

{Given the senator-bill encoding matrix $X$, the first step of standard PCA finds the rank-$1$ matrix that best approximates $X$, i.e.\ it selects unitary vectors $u_1\in\R^m$ and $v_1\in\R^n$ such that the error $\|X-\sigma_1u_1v_1^{\top}\|_F$, evaluated in the Frobenius norm, is minimized. }
Once vectors $u_1$ and $v_1$ have been extracted, PCA is applied again to the deflated matrix in order to extract the second component, and so on; see \cite{calafiore2014optimization}.
PCA can be obtained numerically from the SVD of $X$
\begin{equation}
	X = U \Sigma V^\top = \sum_{i=1}^r \sigma_i u_i v_i^\top,
\end{equation}
where $r$ is the rank of $X$.
For given  $k\in \{1,\ldots,r\}$, we select the first $k$ columns of the  $V$ matrix
$
	V_k := [v_1, \ldots, v_k],
$
and we project the senator-bill data matrix on these $k$ principal directions, obtaining a
projected data matrix $X_k = X V_k\in\R^{m \times k}$.
The choice of how many dimensions to retain is a design parameter and a detailed discussion is postponed to the next section.

\subsubsection{Sparse PCA}\label{ssec:spca} A major issue in PCA is that each of the computed directions {$v_i\in\R^n$ is in general non-sparse. The projected data are thus combinations of all the bills belonging to our dataset.}
 This proves challenging when trying to have interpretability of the directions. In our context we are interested in identifying the bills that play a major role in increasing variance and separating senators. For this goal we apply Sparse PCA.
In the Sparse PCA approach a constraint is added in order to limit the number of non-zero elements in the principal directions \cite{calafiore2014optimization}. More precisely Sparse PCA addresses the problem
of finding unitary vectors $u_1\in\R^m$ and $v_1\in\R^n$ such that
$\|X-\sigma_1u_1v_1^{\top}\|_F$ is minimized subject to the constraint $\|v_1\|_0\leq p$, where $p$ is desired number of non-zero elements in the PCs. The procedure is then iterated for $1\leq k\leq r$.
This approach however is computationally complex for  high dimensions, and algorithms based on convex relaxation to solve it have been proposed in \cite{zou2006sparse,sjostrand2018spasm}.
Similarly to PCA, we then project the senators-bills data matrix on these $k$ PCs obtaining the projected data matrix $X_k = X V_k\in\R^{m \times k}$.

In the next section, a comparative analysis between the two techniques is performed and the main bills extracted by Sparse PCA method are illustrated.

\section{Results}\label{sec:NR}
In this section we compute the Political DNA for all members of the Italian Senate during XVII legislature and present a discussion of the main results.

x
\subsection{Analysis of Political Map in the Italian Senate}\label{subsec:PM}
\begin{figure}[h]
\includegraphics[trim={0cm 0 0 0},clip,width=1\columnwidth]{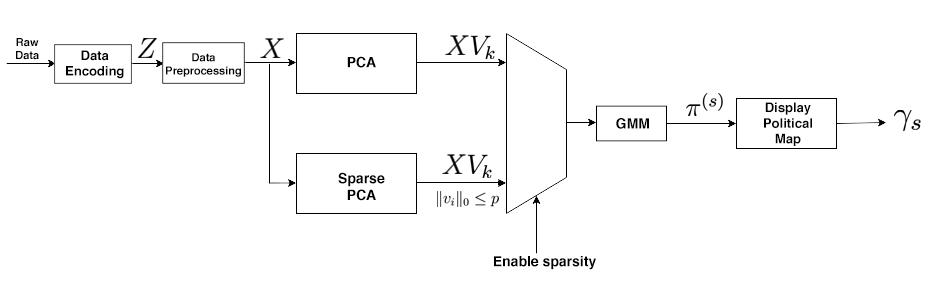}
\caption{Inference procedure for the extraction of Political DNA.
}\label{fig:diagramma}
\end{figure}

The procedure used for the extraction of the Political DNA is summarized in Fig. \ref{fig:diagramma}. The raw data acquired by the dataset are encoded and standardized in the first block, as explained in Section \ref{sub:encoding}. Then, the output is projected in a lower-dimensional space via PCA or Sparse PCA (see Section \ref{sub:DR}). The Gaussian Mixture Model is built using side information on membership extracted from the raw data and the Political DNA is computed via the posterior probability estimation (GMM block, \ref{sec:GMM}).
To produce an interpretable visualization of the results, we represent the political positions of senators in a 2-dimensional space, building what we call a Political Map.

\begin{figure}
\begin{center}
\includegraphics[trim={0cm 0 0 0},clip,width=0.5\columnwidth]{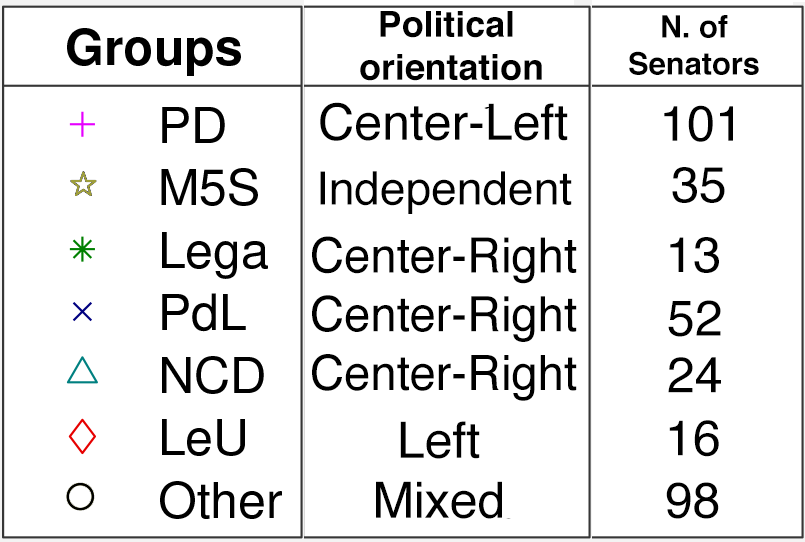}
\end{center}
\caption{Legend of Political Maps: Markers, ideology and number of senators of each group.}\label{fig:legend}
\end{figure}

More precisely, we draw a regular polytope whose vertices represent the political groups. Since the Political DNA is a vector that naturally sums to one, we express the political orientation of each senator as a convex combination of the positions of these vertices.
Formally, given the coordinates for each vertex $\{{a}_{\ell}\}_{\ell\in\{1,\ldots, n_g\}}\in\R^2$, the senator $s$ is identified in the map by a point $\gamma_s\in\R^2$ given by
$$	\gamma_s = \sum_{\ell\in\{1,\ldots, n_g\}} \pi^{(s)}_{\ell} {a}_{\ell}.
$$
The parties are denoted by a different marker (see legend in Fig. \ref{fig:legend}).
It is worth remarking that how to place the groups on the polytope is in principle completely arbitrary.
 Here, parties sharing the same orientation (left-right-independent) are placed on adjacent vertices (see second column in Fig.~\ref{fig:legend}).

\begin{figure}[!t]
\subfloat[][PCA \\ E-Var= $63.38\%$]{\includegraphics[trim={4.55cm 13 3 2},clip,width=0.25\columnwidth]{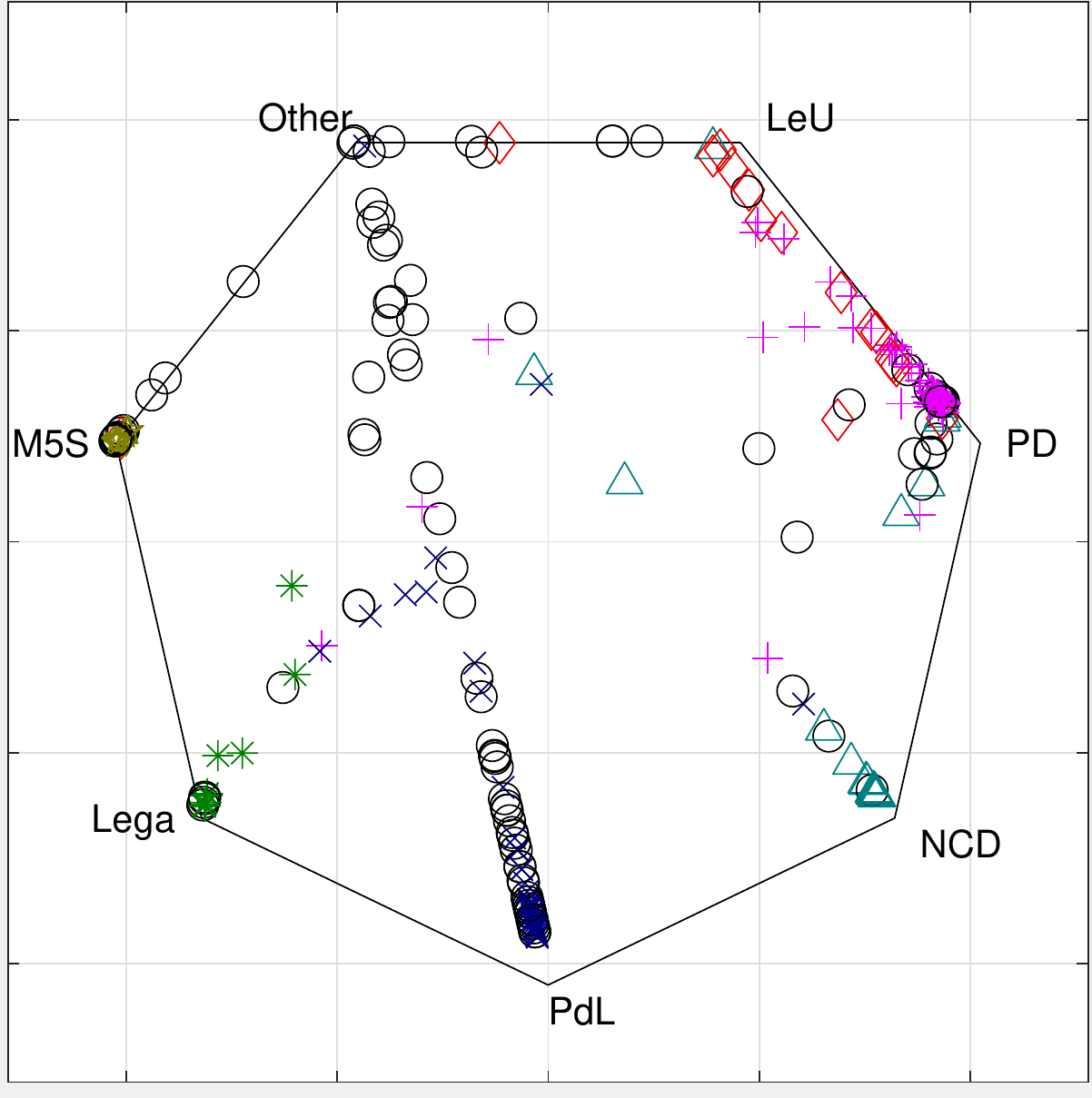}, \protect\label{pca2full}}
\subfloat[][Sparse PCA\\ \centering $p=2$, $ \text{E-Var}=1.33\%$]{\includegraphics[trim={5cm 13 3 2},clip,width=0.25\columnwidth]{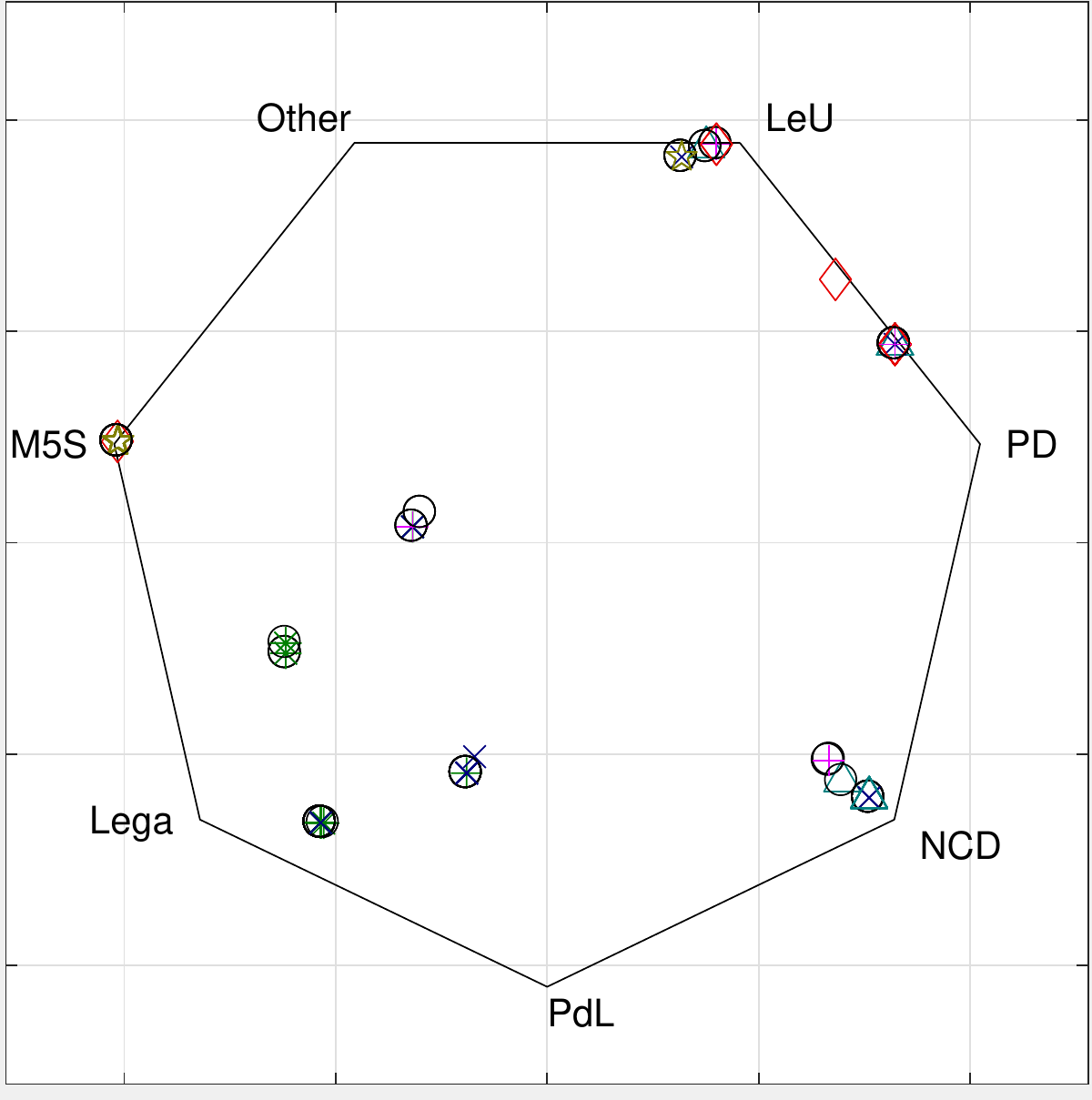}, \protect\label{pca2c2}}
\subfloat[][Sparse PCA\\ \centering$p= 10 $, $ \text{E-Var}=10.01\%$]{\includegraphics[trim={5cm 13 3 2},clip,width=0.25\columnwidth]{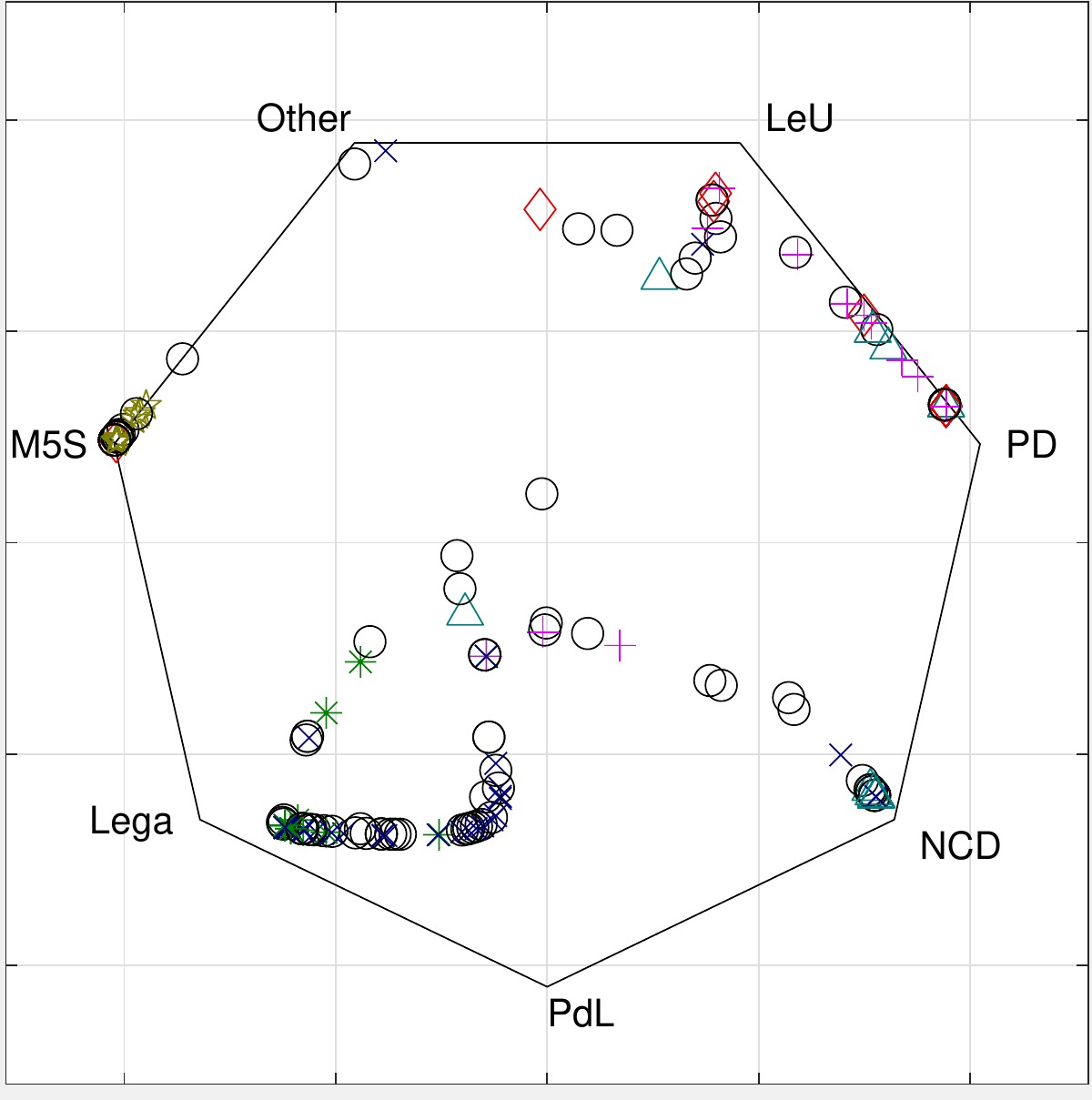}, \protect\label{pca2c10}}
\subfloat[][Sparse PCA\\ \centering$p= 50 $, $ \text{E-Var}=31.26\%$]{\includegraphics[trim={5cm 13 3 2},clip,width=0.25\columnwidth]{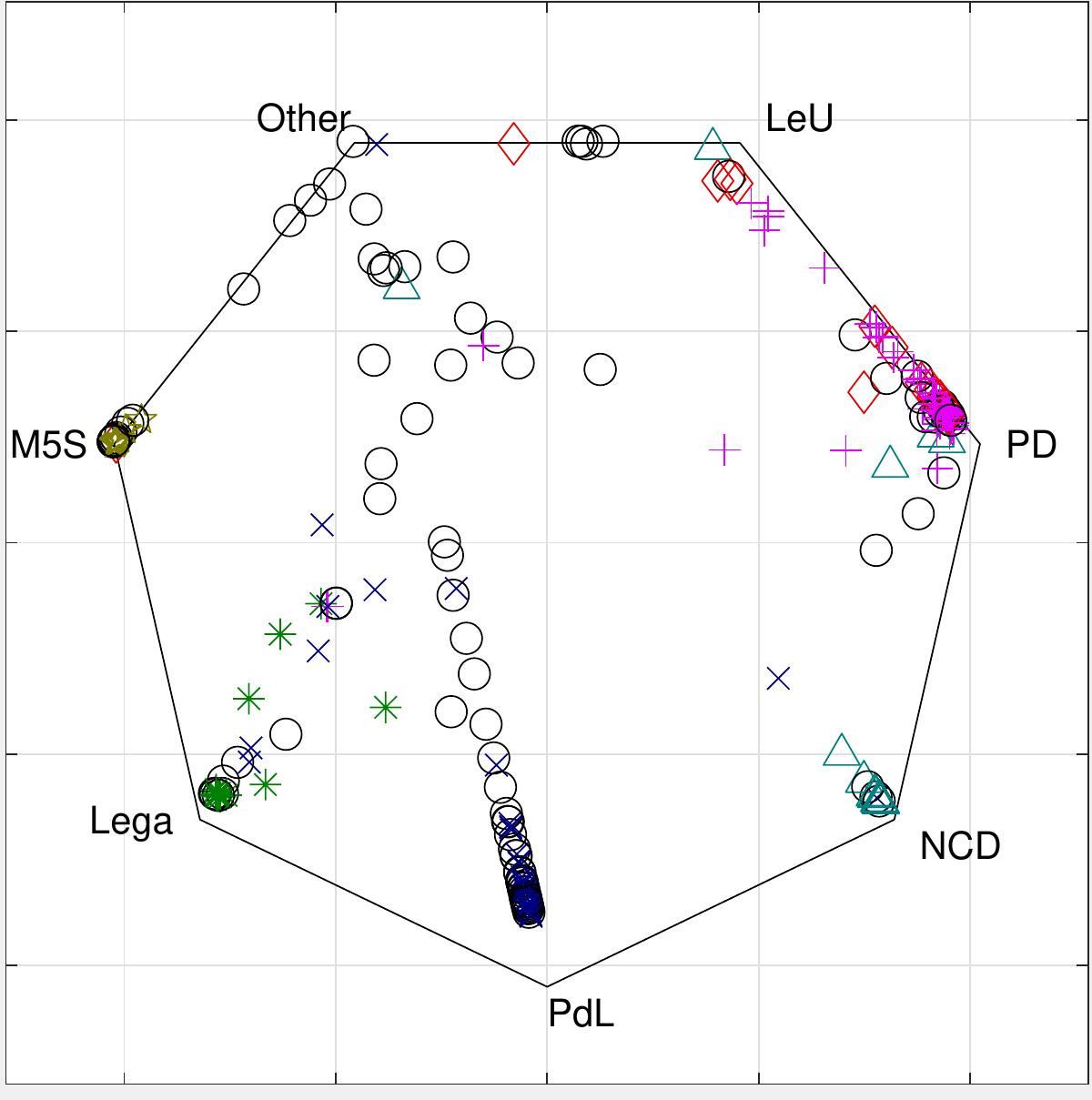}, \protect\label{pca2c50}}
\caption{Comparison of political maps obtained with PCA and Sparse PCA using $k=2$ PCs.}
\label{pca2plot}
\end{figure}

\begin{figure}[H]
\subfloat[][PCA \\ E-Var= $78.79\%$]{\includegraphics[trim={4.55cm 13 3 2},clip,width=0.25\columnwidth]{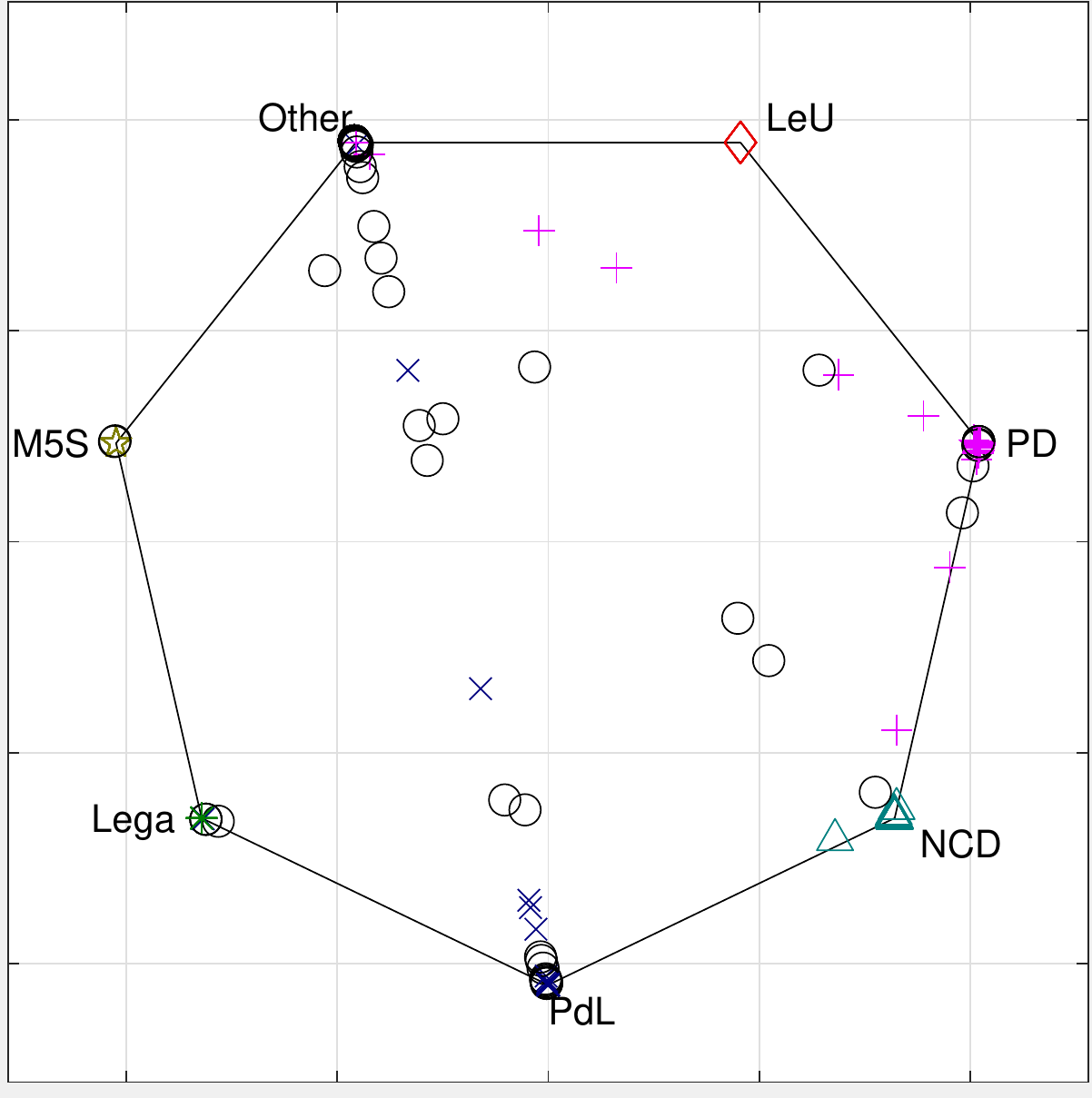}, \protect\label{pca10full}}
\subfloat[][Sparse PCA \centering $p=2$, $ \text{E-Var}=6.14\%$]{\includegraphics[trim={5cm 13 3 2},clip,width=0.25\columnwidth]{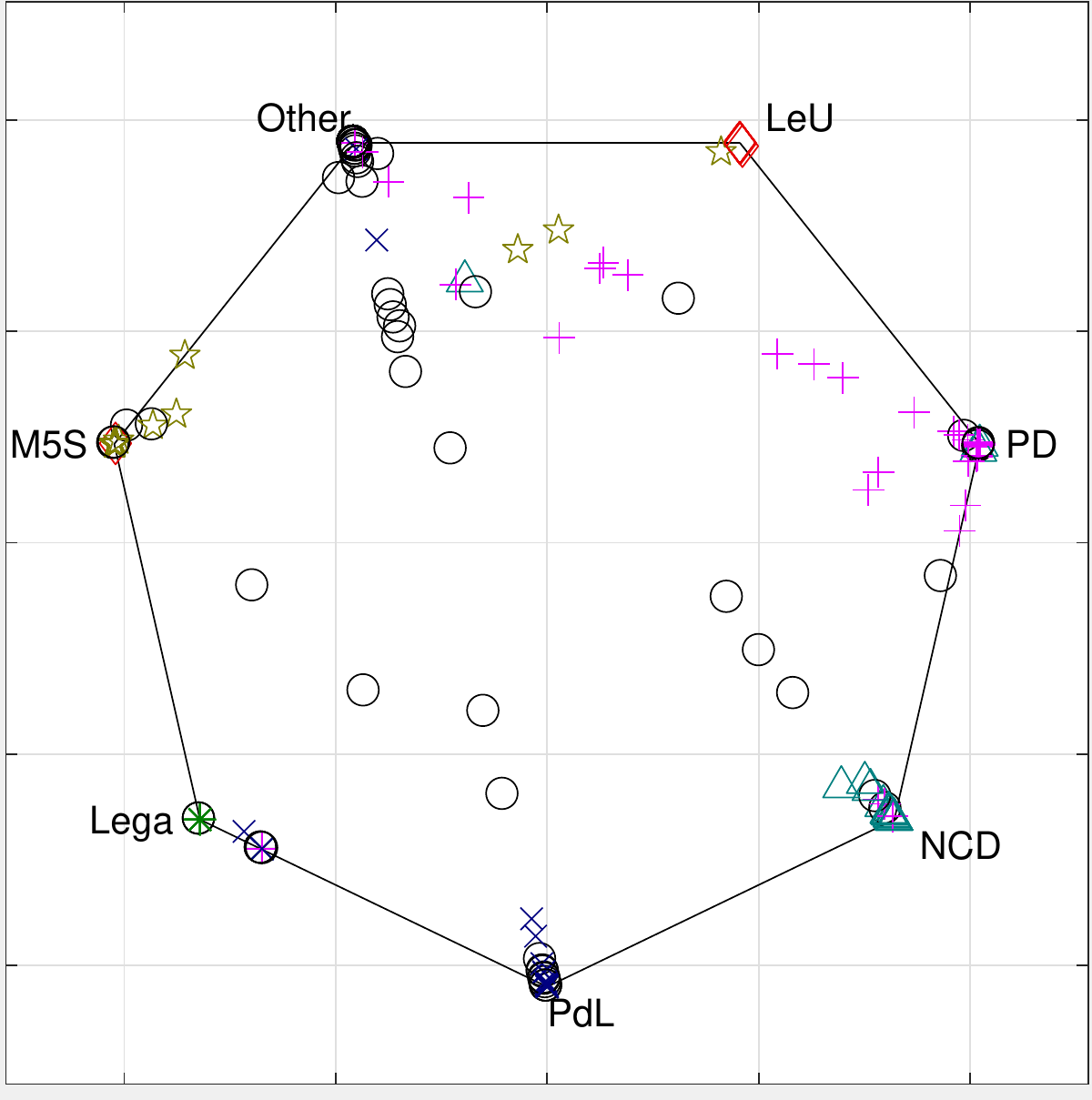}, \protect\label{pca10c2}}
\subfloat[][Sparse PCA \centering $p=10$, $ \text{E-Var}=20.73\%$]{\includegraphics[trim={5cm 13 3 2},clip,width=0.25\columnwidth]{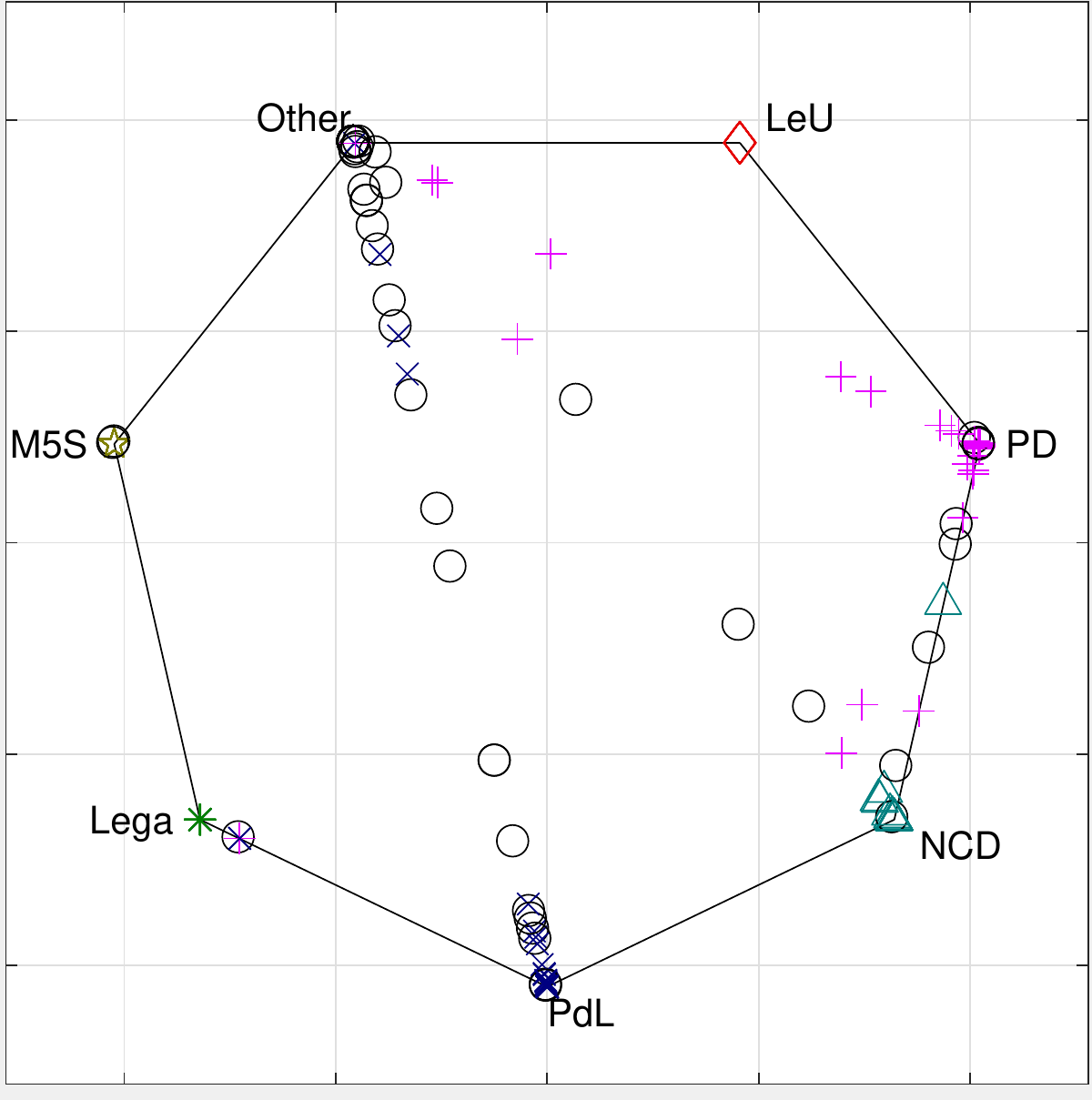}, \protect\label{pca10c10}}
\subfloat[][Sparse PCA \centering $p=50$, $ \text{E-Var}=45.60\%$]{\includegraphics[trim={5cm 13 3 2},clip,width=0.25\columnwidth]{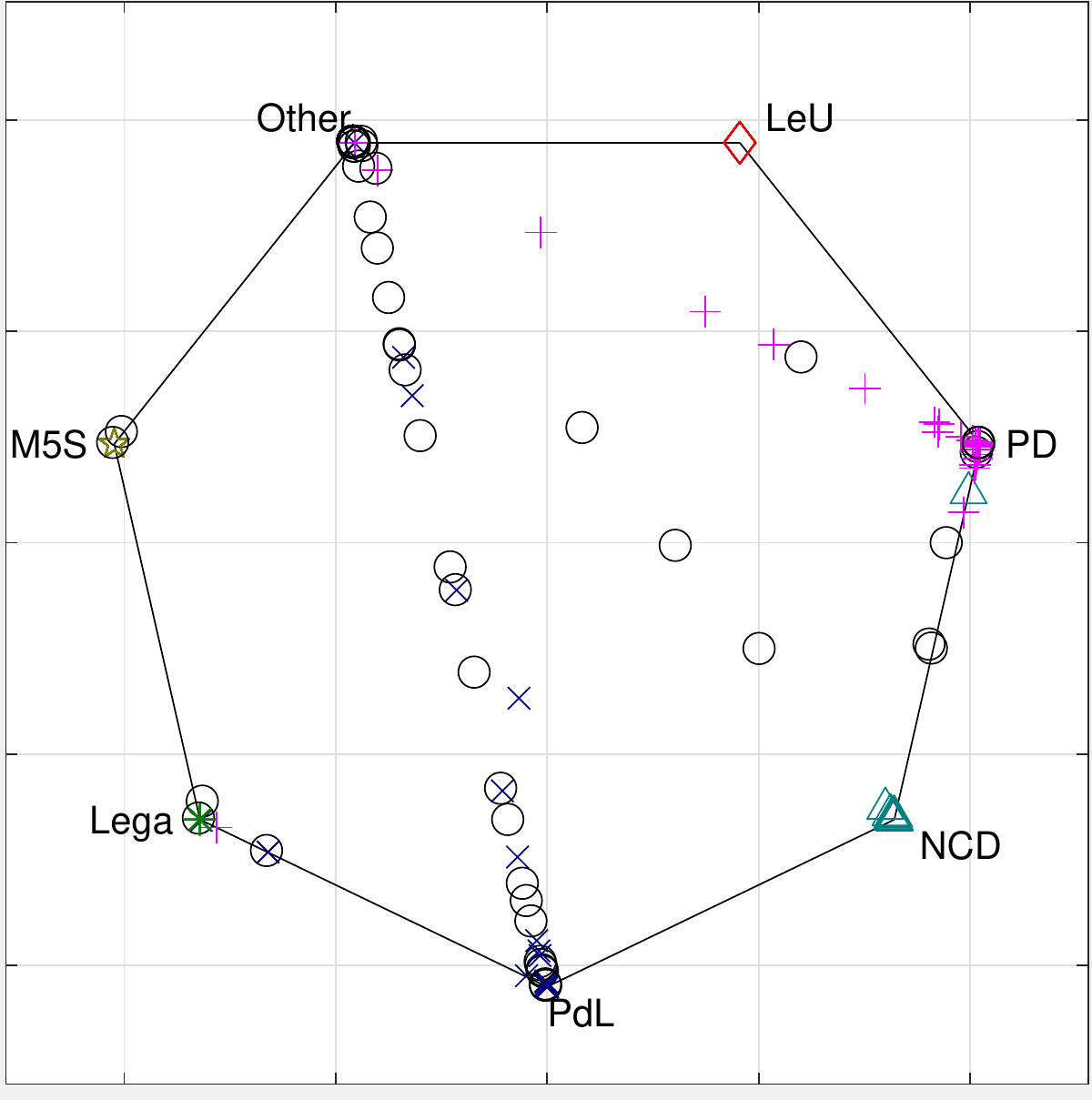}, \protect\label{pca10c50}}
\caption{Comparison of political maps obtained with PCA and Sparse PCA using $k=10$ PCs.}
\label{pca10plot}
\end{figure}

The Political Maps obtained using PCA and Sparse PCA are shown in Fig. \ref{pca2plot} and Fig. \ref{pca10plot}.
More precisely, Fig. \ref{pca2plot}.\subref{pca2full} depicts the results obtained applying PCA and retaining $k = 2$ PCs. Figs. \ref{pca2plot}.\subref{pca2c2}, \ref{pca2plot}.\subref{pca2c10} and \ref{pca2plot}.\subref{pca2c50} are obtained applying Sparse PCA keeping the same number of PCs and enforcing a degree of sparsity equal to $p=2$, $p=10$, and $p=50$, respectively.
 Fig. \ref{pca10plot} shows the comparison of Political Maps obtained applying PCA (Fig. \ref{pca10plot}.\subref{pca10full}) and Sparse PCA (see Figs. \ref{pca10plot}.\subref{pca10c2}, \ref{pca10plot}.\subref{pca10c10}, and \ref{pca10plot}.\subref{pca10c50}) with different level of sparsity and keeping $k = 10$ PCs.

There is no an obvious ground truth to compare against. However, some considerations are in order.
For each experiment, we compute the expressed variance (E-Var), defined as the ratio between the total variance and the variance of data obtained by keeping the first $k$ principal components.
 Looking at the results obtained for $k = 2$ and $k = 10$, we deduce the following facts.
  \begin{itemize}
\item [(i)] As expected, the E-Var of the data increases as a function of the number of PCs and of sparsity level.
\item  [(ii)] Large values of $k$ lead the senators' political positions to shift towards their nominal affiliation, hiding a possible diversification underlying the dataset;
\item [(iii)]  The sparse approach is useful to identify the most significative bills describing the data in the $k$-dimensional subspace. The 10 law proposals identified by the algorithm for the first PCs when $k = 10$ and $p= 10$ are listed in Table~\ref{table:pca10bills}.

\end{itemize}

Regarding the political structure of the Italian Senate:
\begin{itemize}
\item The ``Other'' group is the first one to spread out when projecting the data in a low-dimensional space. This is to be expected, since this group collect senators of very different ideologies, and many senators  ``migrated" to this group after leaving their respective original groups. Hence, the Political DNA allows to recover their original membership or political orientation.
\item The senators belonging to the ``PD'' group, which is the largest one for number of senators and the ruling party during the XVII Legislature, start to shift towards other political parties for large values of $k$ with respect to the other groups, indicating a more heterogeneous internal composition.
\item On the contrary, the ``M5S'' is one of the most compact groups in the Senate, with senators remaining strongly cohesive even for low values of $k$ and low sparsity levels. This can be explained considering that a code of conduct binds the elected candidates to pay a penalty when the guidelines are not followed \cite{Codice_eticoM5S}.
\item For small values of $k$ the ``LeU'' and the ``PD'' senators start to shift towards each other: this is coherent since both groups share a common political orientation (they are both left parties) and the ``LeU'' group foundation is linked to an internal split of the PD group. 
\end{itemize}

\begin{table}[!h]
\centering
\resizebox{0.99\columnwidth}{!}
{%
\begin{tabular}{cc}

\rowcolor{DarkGray} Date  & Description                                                      \\ \hline
 26-11-2013 & Stability Law 2014 - Vote of confidence
                                                                                         \\
\rowcolor{LightGray} 27-11-2013 & Budget Law 2014 - DDL n. 1121. Final vote
                                                                                         \\
05-12-2013 & Decree Extending Military Missions
                                                                                         \\
\rowcolor{LightGray}11-12-2013 & Vote of confidence to the Letta Government
                \\
23-12-2013 & Stability Law 2014 - Vote of confidence
 \\
\rowcolor{LightGray}24-02-2014 & Vote of confidence to the Renzi Government

                                                                                         \\
05-06-2014 &  IRPEF Decree
                                                              \\
\rowcolor{LightGray}08-10-2014 & Enabling act- Jobs Act
                                 \\
19-12-2014 & [Legge di Bilancio 2015] Budget Law 2014 - Bill  n. 1699. Final vote
         \\

\rowcolor{LightGray}25-06-2015 & Bill ''The Good School''
                                     \\
\end{tabular}}
\centering\caption{Bills identified by Sparse  PCA ($k = 10$, $p= 10$).}\label{table:pca10bills}

\end{table}

\subsection{Analysis of Political DNA of outliers}
\label{subsec:P_DNA}
\begin{figure}[!h]
\includegraphics[trim={0cm 0 0 0},clip,width=1\columnwidth]{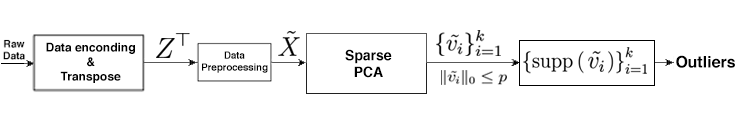}
\caption{Automatic extraction of outliers.
}\label{fig:diagramma2}
\end{figure}

In this section,  we show how Sparse PCA may also be used to automatically select
senators whose voting behavior is not
representative of their nominal group. For these ``outliers," we analyze in depth their Political DNA.
The procedure used for the automatic search of the outlier senators is summarized in Fig. \ref{fig:diagramma2}.

\begin{figure*}[!t]
\centering
\includegraphics[trim={0cm 60 30 70},clip,width=\columnwidth]{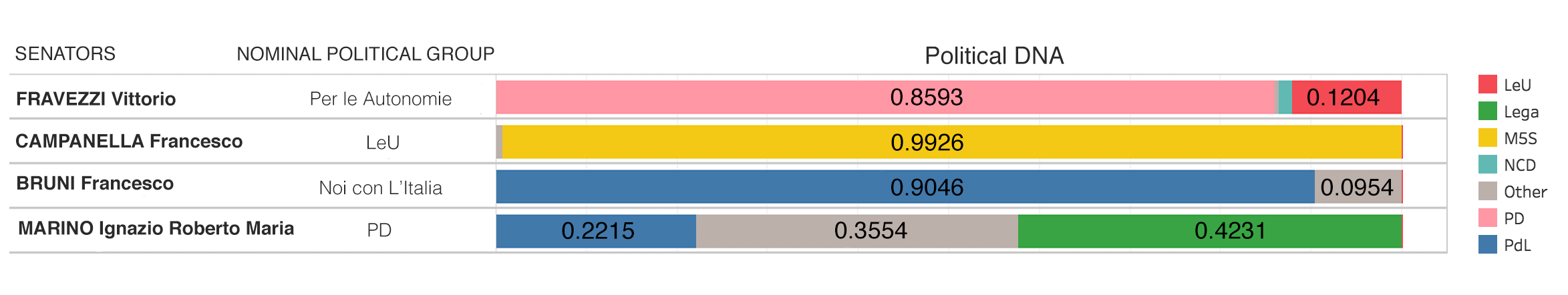}
\caption{Political DNA of  senators extracted via Sparse PCA with $k=2$ and $p=50$}
\label{fig:DNA}
\end{figure*}

The raw data acquired by the dataset are encoded and cleaned. Then the transpose of the vote matrix is standardized, as explained in Section \ref{sub:encoding}. Then we apply the Sparse PCA algorithm (see Section \ref{ssec:spca}) to the transpose matrix, keeping $k=10$ PCs and sparsity degree $p=50$ from which we extract the nonzero components.
The choice of design parameters is motivated by the need of keeping the number of components close to the number of considered groups and sparsity level large enough to have an expressed variance around $40\%$.
In Table \ref{table:2} we show for the 3 PCs the dominant composition, with the corresponding frequency percentage. It should be noticed that the PCs are highly correlated with the clustering in parties.

It is worth remarking that the clustering induced by decomposition obtained with Sparse PCA is robust against changes in the parameter $p$: the composition of PCs remain similar to the ones listed in the table for larger values of~$p$.
Once identified the principal component with party, we isolate the outliers, i.e. senators who are assigned to a principal component/party but who belongs to a different nominal group.

\begin{table}[!h]
\centering
\resizebox{.8\columnwidth}{!}{%
\begin{tabular}{ccc}

\rowcolor{DarkGray} PC  & Most frequent party   & Percentage                                                   \\ \hline
1 & PD & 98\%   \\
 \rowcolor{LightGray} 2 & M5S & 70\%\\
3 & PdL & 68\%\\

%
%
\end{tabular}}\caption{Main features of the first 3 PCs provided by Sparse PCA with $k = 10$ and $p= 10$ ($\text{E-Var}=40.74$).}\label{table:2}

\end{table}

It should be noticed that the parties identified in the 3 first PCs correspond to the 3 major parties present in the Senate during the XVII Legislature. More precisely, the first PC corresponds to the largest political group (PD). In this component just one outlier is present (Fravezzi Vittorio, whose nominal group is in Other).
Looking at the second PC, the one corresponding to M5S, we notice that the most frequent membership of outliers is the Mixed group (66.7\% of outliers are in Mixed Group).  This is coherent with a recent report on the migration of Italian senators during the XVII legislature \cite{Report2017}. Finally, in the third PC (corresponding to PdL) all outliers belong to parties whose political orientation is in the center.
We emphasize that most of the detected outliers are senators who belonged to the party corresponding to the detected PC at beginning of the XVII Legislature and then migrated to other groups.

In Fig. \ref{fig:DNA} we analyze in depth the individual Political DNA of some of outliers provided by the algorithm. The learning is performed by using procedure described in Fig. \ref{fig:diagramma} with $k=2$ and $p=50$ (see also Political Map in Fig. \ref{pca2plot}.\subref{pca2c50}). With these parameters the expressed variance is around 30\% guaranteeing a good compromise between sparsity and data explanations.

\begin{itemize}
\item Fravezzi Vittorio, whose nominal group is Other, appears in the first PC corresponding to PD. He has a strong influence from PD (0.855) and a small influence from LEU (0.136).
\item Campanella Francesco (Nominal group: LeU) appears in the second PC (M5S). This is coherent with the fact that he has been elected with M5S at the beginning of the legislature.
\item Bruni Francesco (Nominal group: NcI (in Other)) is present in the third PC (corresponding to PdL). He has a strong influence from PdL (0.9341) and a small influence from Other (0.0659)
In fact, he was Member of PdL for almost the entire legislature.
\end{itemize}

We report also the Political DNA of Ignazio Roberto Maria Marino, whose nominal group during the entire legislature was PD, but whose Political DNA shows a prevalent component from Lega, and relevant component from PdL.  We refer to \cite{ECC19Plus} for a table containing the Political DNA of all senators.

\section{Conclusions}\label{sec:CR}
We presented an automated numerical technique that, based on publicly available voting data, is able to produce explanatory maps of hidden interconnections among voters nominally belonging to a given number of political or ideological  groups. The method is based on a Gaussian mixture generative model that we use as a prior to compute a voter's posterior influences (the {\em Political DNA}), given evidence of its votes.
We applied our method to a data set pertaining to the votes of 335 members of the Italian Senate on 155 bills during the XVII Legislature.  While the DNA approach is here presented in a political analysis  context, we believe that
the kind of interpretability it offers  makes it suitable to broader application endeavors, such as in the qualitative and quantitative analysis of behaviors, influence and preferences in a marketing context.

\section{Acknowledgment}
The authors are indebted to Vincenzo Smaldore and to OpenPolis Foundation for providing data and  insights on topics related to this paper.

\end{document}